\documentclass[aps,prl,twocolumn,groupedaddress]{revtex4-1}
\usepackage{graphicx}
\usepackage{amsmath}
\usepackage{graphicx}
\usepackage{dcolumn}
\usepackage{bm}
\usepackage{epsfig}
\usepackage{epstopdf}
\usepackage{amsfonts}
\begin{document}
\title{Origins of periodic and chaotic dynamics in microfluidic loop devices}
\author{Jeevan Maddala, Siva A. Vanapalli and Raghunathan Rengaswamy}
\affiliation{Texas Tech University, Lubbock, TX 79401-3121}

\date{\today}

\begin{abstract}
Droplets moving in a microfluidic loop device exhibit both periodic and chaotic behaviors based on the inlet droplet spacing. We propose that the periodic behavior is an outcome of a dispersed phase conservation principle. This conservation principle translates into a droplet spacing conservation equation. Additionally, we define a simple technique to identify periodicity in experimental systems with input scatter. Aperiodic behavior is observed in the transition regions between different periodic behaviors. We propose that the cause for aperiodicity is the synchronization of timing between the droplets entering and leaving the system.  We derive an analytical expression to estimate the occurrence of these transition regions as a function of system parameters. We provide experimental, simulation and analytical results to validate the proposed theory.
\end{abstract}

\pacs{}
\maketitle

Droplets' motion in a microfluidic loop device \cite{ajdari} is a very simple phenomenon that exhibits complex spatiotemporal dynamics; such phenomena are rare in practice \cite{long_periods_natureletters}. A typical microfluidic loop transforms the spacing between the droplets as they travel through the device. It has been shown that a stream of droplets entering the loop at constant inlet spacing could exit the loop at periodic or chaotic intervals \cite{agent,ajdari,sesoms,laypanov_iranian}. The periodic and chaotic behaviors can be observed by changing the inlet spacing of the droplets. While the rules governing the droplets' motion are uncomplicated, discrete decision-making at bifurcations and collective hydrodynamics lead to the intricate dynamics that are observed \cite{fuerstman}. This contrast between the simplicity of the device and the plurality of its response has resulted in this system being investigated by several researchers. In all these studies, the broad questions of interest are: (i) what are the origins of the periodic and chaotic behavior? (ii) is there some common physics that is valid in both the periodic and chaotic regions ? and (iii) is it possible to predict the regions of periodic and chaotic behavior? Although the existing literature partially answers some of these questions such as the prediction of inlet spacings where transitions from periodic to  chaotic behavior occur \cite{sesoms,laypanov_iranian,Passive_trafficking,dynamic_memory}, the cause for these transitions is not well understood.  In this letter, we seek to address the questions posed above as comprehensively as possible at this time. While understanding the loop dynamics is in itself intellectually compelling, the insights gleaned from such an exercise could also help in understanding natural systems such as blood flow in microvasculature and in developing practical design and control approaches in the field of discrete droplet microfluidics.

\begin{figure}[h]
 \includegraphics[height = 2.5 cm, width=0.45\textwidth]{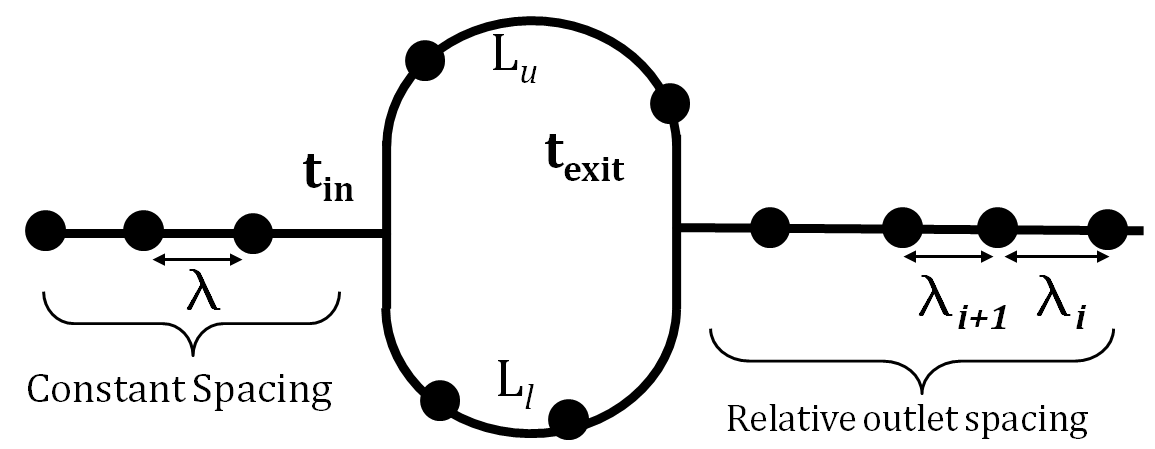}%
 \caption{A constant stream of droplets passing through a microfluidic loop device}
 \label{fig:Loop_represenation}
 \end{figure}

Consider a train of droplets entering a microfluidic loop device as shown in Fig \ref{fig:Loop_represenation} at a constant inlet spacing $\lambda $.  The relative spacings between the droplets remain constant until they reach the loop junction. As the droplets enter the upper or lower branch, the resistance of that branch changes. This results in a redistribution of the bulk flows to the upper and lower branches. Since the droplets' velocities are proportional to the bulk flow, the droplets in the upper and lower branches move at different velocities leading to changes in the relative distances between them as they exit the loop. Mathematically, the loop device $(G)$ transforms the inlet spacing $(\lambda)$ to a sequence, $\{\bar{\lambda}\}_n = \{G(\lambda)\}_n$. It has been shown in the literature that the exit spacings can either form a periodic or a chaotic pattern. These patterns are captured through a Poinc\'are map, where the number of distinct clusters represent the number of periods. There are two disadvantages to such a characterization: it is difficult to exactly identify the number of periods when confronted with scatter in the data, and the plot does not explain the physics of the process that leads to these periodic or aperiodic behaviors. To address these two issues, we start with a mathematical definition of periodicity.  The exit sequence $(\{\bar{\lambda}\}_n)$ is periodic with period $p$, where $p$ is the smallest integer such that $\{\bar{\lambda}\}_i = \{\bar{\lambda}\}_{i+p}$ $\forall$ $i \;\epsilon\; \mathbb{N}$. The exit sequence is aperiodic, if there exists no $p$ such that $\{\bar{\lambda}\}_i = \{\bar{\lambda}\}_{i+p}$ $\forall$ $i \;\epsilon\; \mathbb{N}$. From the definition of periodicity it can be noticed that the periodicity condition will hold for all integer multiples of $p$. Further, it is self evident that the periodicity condition cannot hold for two $p$'s that are not multiples of each other. However, if the periodicity condition is valid in some region, {\em i.e.}, $ i \; \epsilon \; (N_1,N_2)$ for some $p$, then that behavior is termed locally periodic. We are interested in the physics that leads to periodic behavior and the manner in which the system transitions from periodic to chaotic behavior and {\em vice versa}.

\begin{figure}[h]
 \includegraphics[height = 4cm, width=0.45\textwidth]{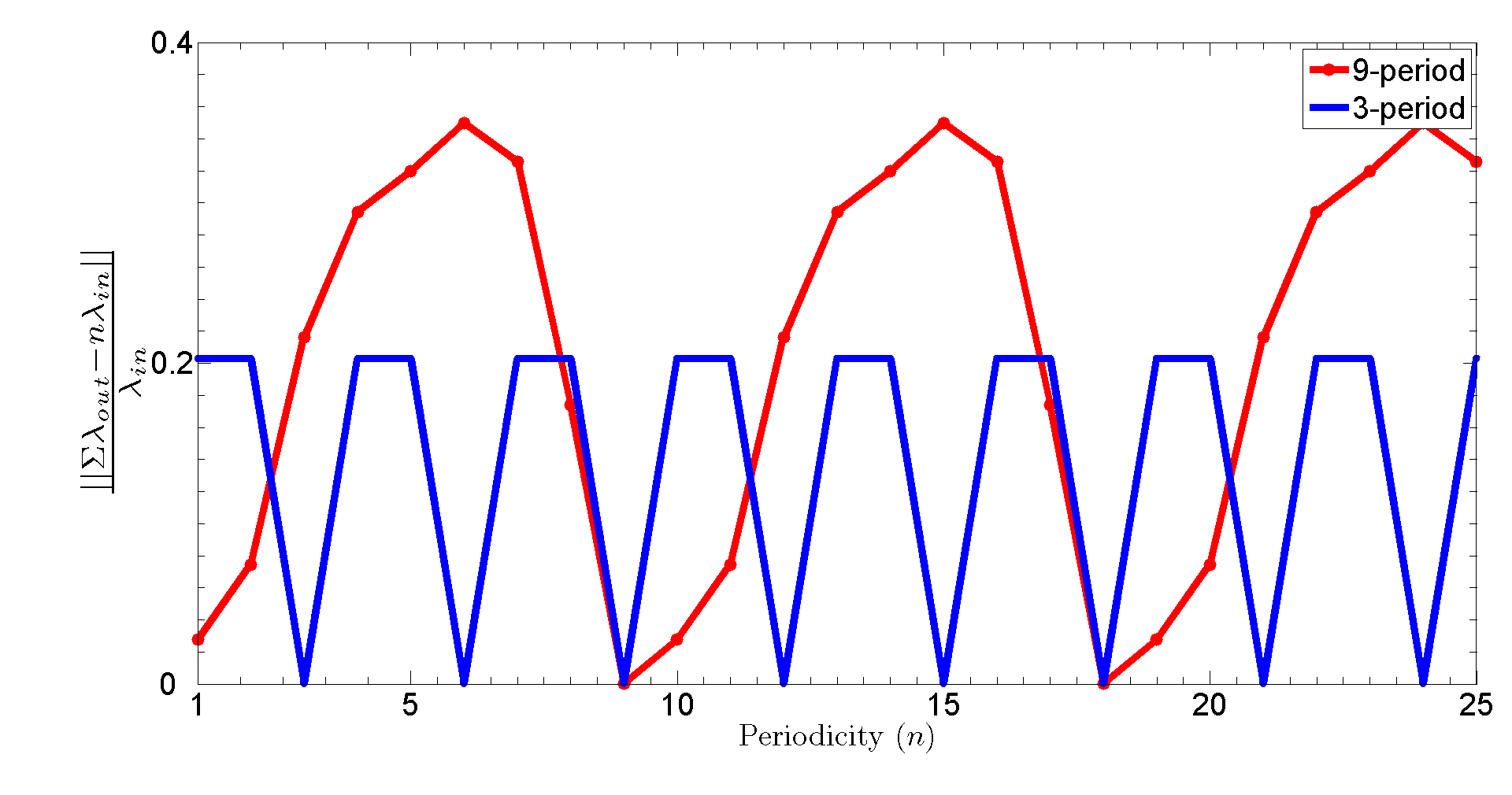}%
 \caption{Conservation of droplet spacing in the case of $3$ and $9$ period behaviors}
 \label{fig:Conservation}
 \end{figure}

The sustained periodic behavior is a result of a series of droplets taking the same decisions repeatedly - this can also be viewed as memory of the system \cite{dynamic_memory}. The decisions will be repeated with period $p$ if the $i^{th}$ and $i+p^{th}$ droplets undergo the same experience \textit{i.e.,} velocity changes as they pass through the loop. This happens if the loop snapshot (by which we mean the device with droplets at particular positions in the loop) that the $i+p^{th}$ droplet encounters is the same as the one encountered by the $i^{th}$ droplet. A natural consequence of this concept is that the amount of the bulk phase fluid trapped between the $i^{th}$ and $i+p^{th}$ droplets should exactly equal the amount of the fluid trapped between the same droplets prior to entering the loop. Mathematically, the bulk fluid present in between  $i^{th}$ and $i+p^{th}$ droplets prior to the device entry is $p \lambda S$, where $S$ is the cross sectional area of the channel. Assuming bulk phase incompressibility, the volume trapped between the droplets after exit is the sum of all the exit relative spacings multiplied by $S$. This volume conservation that is required for repetition translates to a remarkable ``conservation of droplet spacing" given below:
\begin{eqnarray}
&&\bar{\lambda}_{i+1} + \bar{\lambda}_{i+2} + ... + \bar{\lambda}_{i+p} = p \lambda \nonumber \\
&&\lambda = \frac{\bar{\lambda}_{i+1} + \bar{\lambda}_{i+2} + ... + \bar{\lambda}_{i+p}}{p}
\label{Eqn:consrvtn}
\end{eqnarray}
If there exists a unique $p$, $\forall$ $i$ $\epsilon$ $\mathbb{N}$ then eqn \ref{Eqn:consrvtn} is an \textit{`if and only if'} condition for a loop device to show sustained periodic oscillations. We validate this `conservation principle' that is derived through physical arguments using the network model proposed in \cite{ajdari}. We model two experimental scenarios, 3-period and 9-period behaviors, without input fluctuations. We plot the value of  $|\Sigma_{k=l}^{l+j-1} \bar{\lambda} - p \lambda|/\lambda $  as a function of $j$ as shown in Fig \ref{fig:Conservation}. If our insight is correct then this value should be equal to zero whenever $j$ is equal to multiples of $p$, the periodicity of the system. The initial $(l)$ that is used in the equation does not have any impact on the result because the case studies considered here have sustained periodic oscillations. As expected, in Fig \ref{fig:Conservation}, the curve touches the x-axis at $N=3,6,9 \hdots$ for the 3-period case and
$N = 9,18 \hdots$ for the 9-period case, validating the `conservation principle'.

\begin{figure}[h]
 \includegraphics[height = 5cm, width=0.45\textwidth]{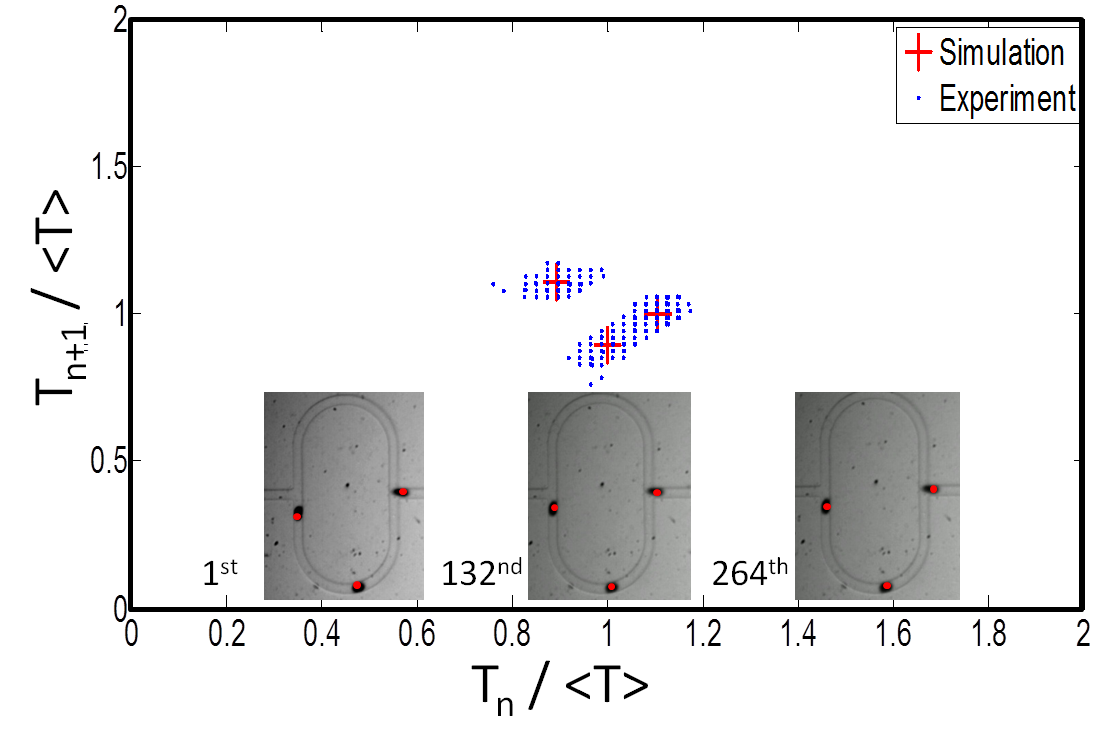}%
 \caption{Three droplets exit the loop device in between the images $1^{st}$, $132^{nd}$ and  $264^{th}$. Poincare map of the exit times shows a two period behavior, whereas the simulation shows three periods. Oil and water flow rates are $550$ and $25$ $\mu$ L/hr respectively}
 \label{fig:Snap_shotcomp}
 \end{figure}

Estimation of the period $p$ in case of experimental systems can be complicated due to fluctuations in the inlet spacing. In such cases using conventional methods such as Poinc\'are maps \cite{fuerstman} to obtain periodicity may be misleading. The conservation principle provides another view of periodicity. Looking at eqn \ref{Eqn:consrvtn}, it is apparent that the periodicity is equal to the number of droplets that exit before a snapshot repeats. Consider a 2-period behavior as identified by a Poinc\'are map depicted in Fig \ref{fig:Snap_shotcomp}. Modeling the system without experimental fluctuations at the inlet using a network model leads to a 3-period behavior. To confirm if this is the true period of the system we used the snapshot comparison technique derived from the conservation principle. The $1^{st}$ image is repeated after the $132^{nd}$ and $264^{th}$ snapshots as shown in the inset of Fig \ref{fig:Snap_shotcomp}, with the droplets exiting at the $1^{st}$, $132^{nd}$ and $264^{th}$ images. In this case, three droplets exit before the original image is repeated, confirming that this indeed is 3-period behavior, contrary to what Poinc\'are map suggests. The proposed snapshot-comparison method is therefore particularly helpful in estimating the periodicity of loop dynamics where experimental scatter might otherwise make such identification problematic.\\

\begin{figure}[h]
 \includegraphics[height = 4.5cm, width=0.45\textwidth]{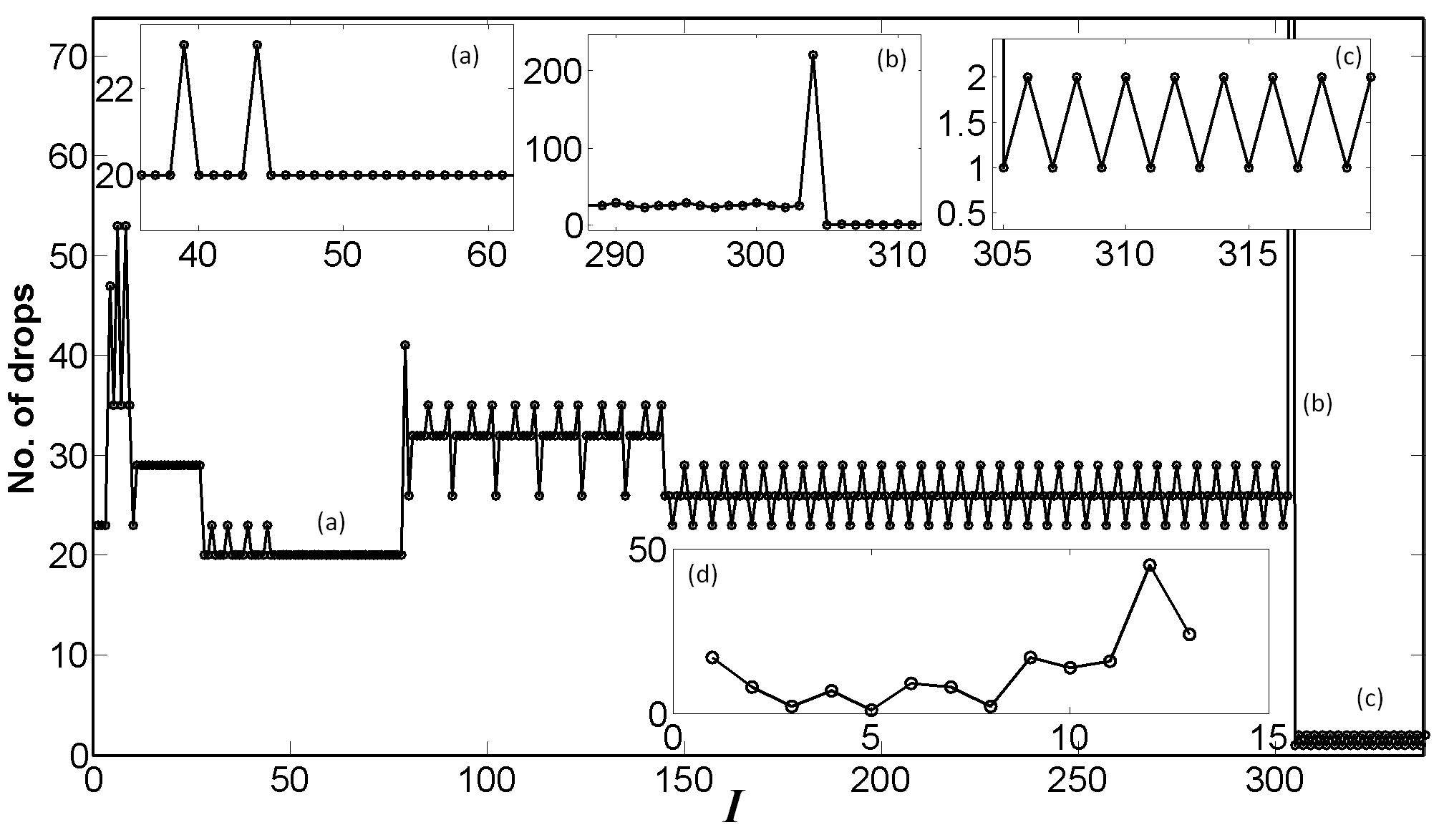}%
 \caption{Periodicity calculated using Eqn \ref{Eqn:consrvtn} for an aperiodic system at distinct intervals of time index; (a),(b),(c) show distinct regions of the above plot, inset (d) shows an aperiodic experimental system demonstrating conservation of spacing. Oil and water flow rates are $250$ and $25$ $\mu$ L/hr respectively}
 \label{fig:Intermmittency}
 \end{figure}

We now turn our attention to the understanding of aperiodic behavior and the physics that leads to the transitions from periodic to aperiodic behavior and {\em vice versa}. We answer this transition question by exploring what happens to the conservation principle in the aperiodic region. There are two possibilities: one in which the conservation principle is never valid for any two pairs of droplets and the other scenario is where the conservation principle is valid between some droplet pairs but not repeatedly {\em ad infinitum}. To resolve between the two possibilities we again take recourse to the the network model \cite{ajdari} where we simulate a microfluidic loop for aperiodic behavior. In Fig \ref{fig:Intermmittency}, the x-axis of the plot is increasing natural numbers ($I$) representing the $I^{th}$ time that the conservation principle holds and the y-axis representing the number of the droplets that participate in the conservation equation (the measure of periodicity) when it holds for the $I^{th}$ time. This plot suggests that even in the aperiodic case, the conservation principle is still driving the physics; however, the periodicity changes with time and the conservation equation is valid at all these periodicities. Inset Fig \ref{fig:Intermmittency}(a) shows the initial period, where for a while a local periodic behavior is observed, which is broken, followed by the same period for a shorter time and so on. This is a classic case of intermittent behavior; similar behavior is observed in inset Fig \ref{fig:Intermmittency}(b). Inset Fig \ref{fig:Intermmittency}(c) shows the dynamics settling down to a three period behavior for a larger duration but the conservation principle holds for alternate 1 and 2 droplets and, as a result for 3 droplets at all times, demonstrating the richness of the behavior exhibited by this simple loop device. Fig \ref{fig:Intermmittency}(d) shows an experimental system displaying aperiodic behavior where the conservation principle is valid for distinct sets of droplets. This result has implications on how the system transitions from periodic to aperiodic behaviors and suggests that the aperiodic behavior is likely to be interspersed between the periodic regimes.

\begin{figure}[h]
 \includegraphics[height = 3 cm, width=0.45\textwidth]{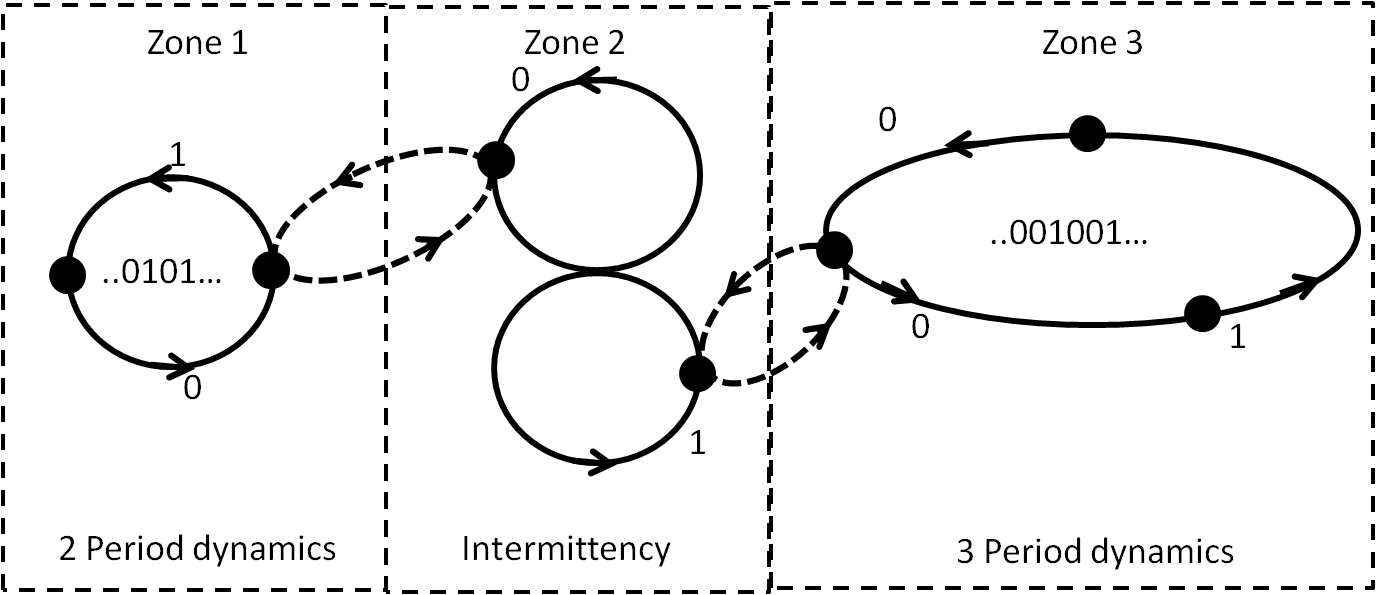}%
 \caption{Representation of transition phenomena in a microfluidic loop device}
 \label{fig:Cellr_automat_pic}
 \end{figure}

\begin{figure}[h]
 \includegraphics[height = 5cm, width=0.45\textwidth]{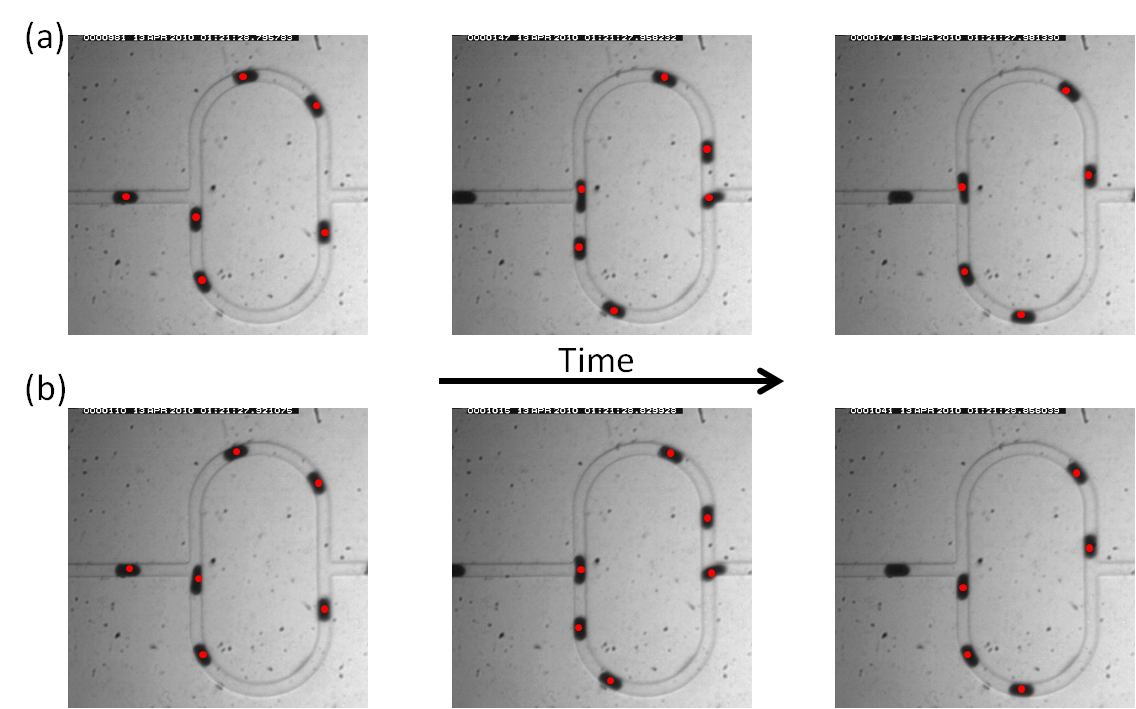}%
 \caption{A loop device showing transitions: (a) A sequence of snapshots showing the entering droplet choosing the upper branch; (b) At a later time in the same experiment, with similar conditions, the entering droplet chooses the lower branch. Oil and water flow rate are $250$ and $25$ $\mu$ L/hr respectively }
 \label{fig:Exp_comp}
 \end{figure}

The intermittent transitions between periodic behaviors in an aperiodic scenario is depicted in Fig \ref{fig:Cellr_automat_pic}, where the values `0' and `1' represent the binary decisions of the droplets \textit{i.e.,} choice of either the upper or the lower branch. In the transition regions any combination of $0$'s \& $1$'s is possible. This can result in large intermittent periods and sporadic transitions to other periodic (for example 2 or 3) behaviors. It can be noticed that a remarkably similar picture is observed in the intermittency graph derived from the conservation principle as shown in Fig \ref{fig:Intermmittency}. Interestingly, similar phenomena are also observed in cellular automata \cite{cellaratmta}. Fig \ref{fig:Exp_comp} shows an experimental scenario where the intermittent behavior is observed. Consider Fig \ref{fig:Exp_comp}(a); the snapshots are arranged in an increasing order of time. In all these images the droplets enter the loop when the loop contains 5 droplets.  As the exiting drop leaves, the entering drop can either take the lower or the upper branch. In the case of Fig \ref{fig:Exp_comp}(a), the entering droplet chooses the upper branch, but in the case of Fig \ref{fig:Exp_comp}(b), the droplet chooses the lower branch. Although the system has very similar conditions at both times, the dynamics of this system moves to a completely different regime due to a single droplet altering its decision. Since such transitions happen at different time instants during the course of the droplets' travel through the loop, the dynamics of the system becomes unpredictable.

In summary, the transition region is where the conservation principle holds for several periods at different times; if the droplets take decisions in a repetitive fashion {\em ad infinitum}, then the behavior is periodic. For aperiodic behavior to be manifested, there has to be a critical event to break the constant periodic cycle. We propose that this critical event is one where the droplets enter and exit at around the same time, leading to changes in the droplets' decisions. This will in turn change the amount of dispersed phase fluid trapped between the droplets. Predicting the onset of transitions is of significant importance. Reliable prediction of the transition region can be used to design experiments and to manufacture devices that will consistently show either periodic or chaotic behavior. It is clear that there are an infinite set of $\lambda$ values in probably a very small range that characterizes the transition region. Predicting the range of $\lambda$ values where the transitions occur is a difficult problem. Instead, in this work, we attempt to predict a particular value of $\lambda$ ($\lambda_c$) in the transition region. From a practical standpoint, since we expect the transition regions to be characterized by a very small range of $\lambda$ values, prediction of one value in the range should suffice. To predict one $\lambda_{c}$, we need to find a $\lambda$ that results in a scenario where a droplet enters the loop device while another droplet is exiting. While this could happen for several different pairs of droplets (leading to a range of $\lambda$ values), we assume that the critical $\lambda_c$ is when the first droplet exits the loop device while another droplet enters the device. However, depending on the $\lambda_c$, the number of droplets that a loop holds ($n$) before the first droplet exits could be different. Remarkably, we will demonstrate that by changing $n$, all the transition regions can be identified. Interestingly, the $n$ used in the calculation of a particular transition $\lambda_c$ does not provide any indication of the period $p$ of the sustained periodic behavior that follows the transition. This is because the transition $\lambda_c$ has no bearing on the number of droplets for which the conservation principle holds. Consider Fig.\ref{fig:Loop_represenation} - the entering droplets move at a velocity $v = \beta Q/S$, where $\beta$ is the slip factor, $Q$, $S$, $L_l$ and $L_u$ are inlet flow rate, cross sectional area of the channel and length of upper and lower branches respectively. Assuming that the loop is symmetric ($L_l \approx L_u$), droplets alternate between top and bottom branches. The resistance of the lower branch is $R = R_l + m R_d$, where $R_l$ is the branch resistance, $R_d$ is the droplet resistance and $m$ is the number of droplets in the branch.

The exit time of the first droplet after $n$ droplets enter the loop device is as follows:
 \begin{eqnarray}
t_{1,exit} = \frac{L_l- \sum_2^n \lambda \left(\frac{R_u+floor((k-1)/2)R_d}{R_u+R_l+(k-1)R_d}\right)}{\beta Q/S \left(\frac{R_u+R_d}{R_u+R_l+kR_d}\right)}
\end{eqnarray}
In the above equation, $t_{1,exit}$ is the time taken by the droplet from that instant to exit the loop, as shown in Fig \ref{fig:Loop_represenation}, and is not the total time it takes to exit the loop. The entry time of the next droplet in the loop is given by $t_{entry} = \frac{\lambda S}{\beta Q}$. Therefore the critical ($\lambda_c$) values occur when: $t_{1,exit} = t_{entry}$. This results in (with $\bar{R} = R_d/R_u$)
\begin{figure}[ht]
 \includegraphics[height = 4cm, width=0.5\textwidth]{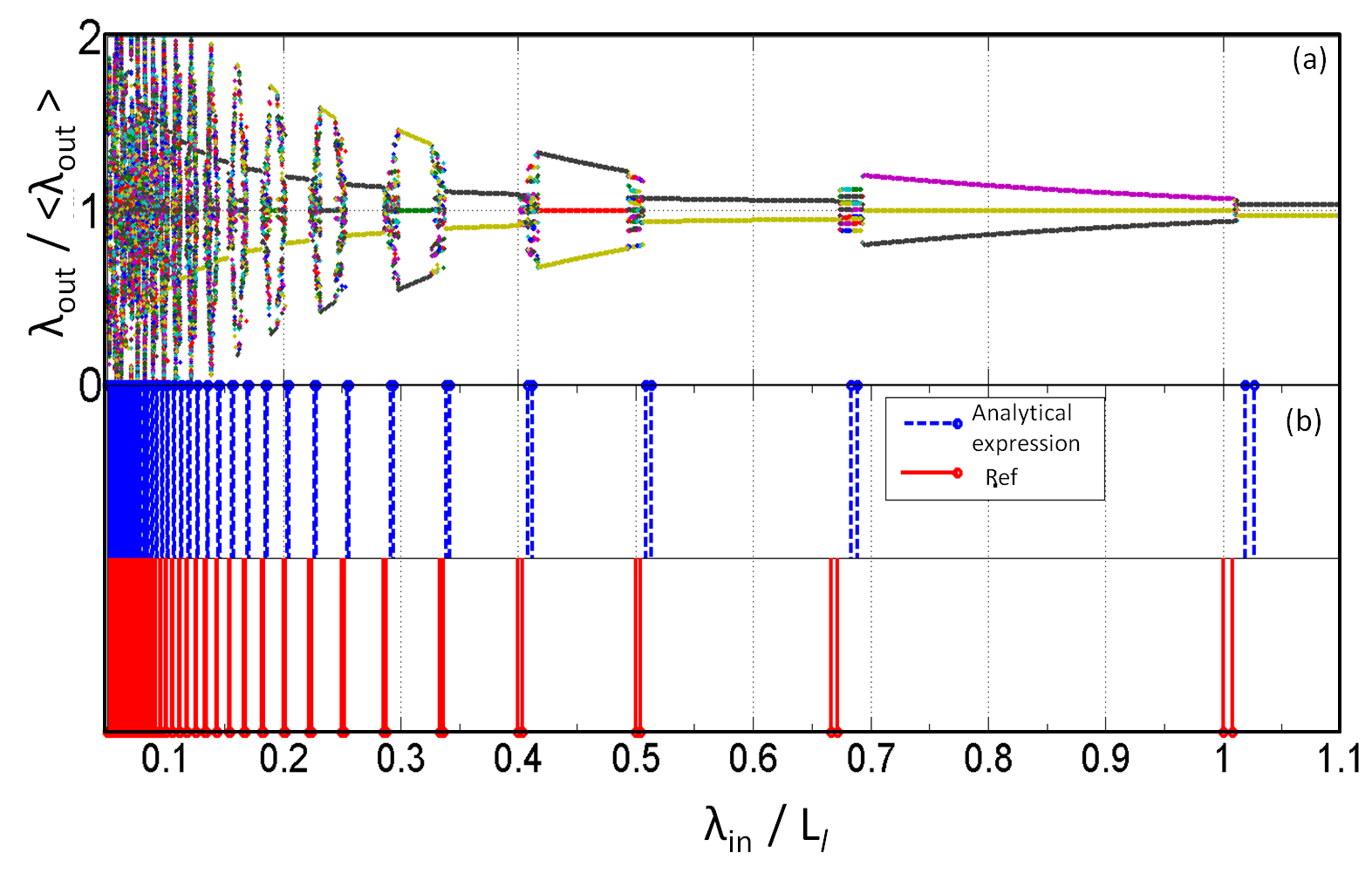}%
 \caption{Bifurcation map simulated by changing the input spacing in a scenario where the oil and water flow rates are $550$ and $25$ $\mu$ L/hr respectively, the bottom plot shows the prediction of transitions in the bifurcation maps using the proposed analytical expression along with that reported in the literature\cite{sesoms}. The upper and lower branch lengths of the loop device are $2235 \mu m$ and $2217 \mu m$, respectively }
 \label{fig:bifurcation}
 \end{figure}

 \begin{eqnarray}
\frac{L_l}{\lambda_{c}} &=& \sum_{k = 1}^n \frac{R_u}{R_u+R_l+k R_d} + \sum_{k = 1}^n \frac{floor(k/2)R_d}{R_u+R_l+kR_d}   \nonumber \\
\frac{L_l}{\lambda_{c}} &=& \sum_{k = 1}^n \frac{1+floor(k/2)\bar{R}}{2+k \bar{R}} \;\;\; if\;\;\; R_l\;\; \approx\;\; R_u
\label{eqn:analytical}
\end{eqnarray}
The critical $\lambda_c$ values are obtained by varying $n, n \in \mathbb{N}$. The proposed analytical expression is validated using a bifurcation map along with an analytical expression from the literature \cite{sesoms}.

A particular loop configuration is simulated to validate the theoretical predictions as shown in Fig \ref{fig:bifurcation}. The flow rate is kept constant, and only the feeding frequency is varied. Based on input frequency, after the initial transients, the droplet exit pattern reaches a steady state. At this point, periodicity is calculated and plotted against the inlet spacing. Fig. \ref{fig:bifurcation}(a) shows a bifurcation map, where normalized inlet spacing is plotted on the $X$-axis and the normalized outlet spacing is plotted on the y-axis.

At $\lambda > 1$, the system shows a steady $2$-period behavior, but as the spacing decreases, the outlet periodicity transitions from a 2-period to a $3$-period behavior - observe that there is no period doubling \textit{vis a vis} logistic map \cite{strogatz}. The 3-period behavior lasts for some time and the system goes back to a two period behavior. After the next transition, the system drifts to a 4-period behavior. We observe only $3$ values repeating within the 4-period pattern. The analytical expression captures all these transition regions corroborating the intuition that droplets entering and exiting at the same time leads to chaotic behavior. The results from the expression for critical $\lambda_c$ values proposed in the literature are also plotted in Fig. \ref{fig:bifurcation}(b) and are in agreement with the theory that is proposed in this paper.

In conclusion, in this paper, we propose a conservation principle that is valid for periodic and chaotic behavior in microfluidic loop systems. Our experimental and theoretical analysis provides a unique way to quantify this behavior. As the conservation of spacing is valid for both periodic and chaotic regions of the loop device, this device resembles a Hamiltonian system\cite{hamiltonian_Loop}. The chaotic behavior is a result of transitions occurring when droplets enter and exit at the same time. This is akin to complex sequences derived with simple rules in Cellular Automata \cite{cellaratmta}. We derive an analytical expression for predicting the transition regions based on the new physical insights proposed in this paper. The results suggest that the system achieves chaotic behavior through a phenomenon called ``intermittency route to chaos''\cite{strogatz,Ott} and not just because of a 3 period behavior \cite{threeperd_chaos}.

We thank Prof. Jerzy Blawzdziewicz, Dr. Zeina S. Khan and William S. Wang for valuable discussions. We also thank Dr. Babji Srinivasan and Swastika S. Bithi for experimental support. We acknowledge financial support from National Science Foundation (Grant No. CDI-1124814).

\bibliography{apstemplate}
\bibliographystyle{apsrev4-1}
\end{document}